# AI Civilization and the Transformation of Work

*AI 문명 시대와 일자리*

Dongsoo Han, School of Computing, KAIST

## Abstract

The emergence of artificial intelligence (AI) and robotics is catalyzing a profound transformation in the nature of human labor, fueling a contentious debate about the future of employment. While prominent studies predict substantial job displacement due to automation, historical precedents from past technological revolutions suggest that innovation tends to expand, rather than shrink, the scope of economic activity and employment in the long run. This paper advances the thesis that the transition to an "AI Civilization" will fundamentally restructure the mechanisms of employment creation. We argue for a paradigm shift from a centralized model—where a limited number of organizations create jobs for the masses—to a decentralized ecosystem where individuals are empowered to generate their own employment opportunities. This shift is enabled by AI-driven productivity augmentation, which dramatically lowers the barriers to creating economic value. Drawing on an analysis of economic history, contemporary data on labor market dynamics, and the growth of digital platforms, this paper posits that human-AI co-evolution will significantly increase individual productivity and open new frontiers of economic activity. We explore the implications of this structural transformation for education and workforce development, concluding that the focus must shift from rote knowledge accumulation to cultivating skills in human-AI collaboration, creative problem-solving, and the design of novel economic domains. This paper contributes to the literature by offering a forward-looking framework that emphasizes the decentralizing potential of AI on labor markets, moving beyond the traditional displacement-versus-creation dichotomy.

**Keywords:** AI civilization, automation, future of work, technological unemployment, human–AI collaboration, decentralized employment, creator economy

# 1. Introduction

The rapid proliferation of artificial intelligence (AI) has positioned humanity at the cusp of a new socio-economic era, often termed the "AI Civilization." This transition, characterized by unprecedented advances in machine learning, robotics, and data processing, is fundamentally reshaping the global economy and the very nature of work [1, 2]. Recent developments in generative AI have intensified a long-standing debate, with reports from institutions like Goldman Sachs suggesting that the equivalent of 300 million full-time jobs could be exposed to automation [3]. Similarly, the World Economic Forum's Future of Jobs Report 2023 projects that nearly a quarter of all jobs will be disrupted in the next five years, with 83 million roles eliminated and 69 million new ones created, resulting in a net decrease of 14 million jobs [4]. Such forecasts have amplified societal anxieties about mass technological unemployment.

Historically, however, technological revolutions have consistently elicited such fears. The mechanization of agriculture during the Industrial Revolution and the computerization of clerical work in the 20th century both triggered widespread concern over job displacement [5]. Yet, long-term economic history reveals a consistent pattern: technological progress ultimately reshapes rather than eliminates labor markets, creating new tasks and industries that expand the frontier of economic activity [6, 7]. The central question is not whether AI will destroy jobs, but how it will transform them and what new opportunities will emerge.

This paper argues that the AI civilization marks a pivotal departure from previous technological epochs by fundamentally altering the structure of employment creation itself. We move beyond the conventional dichotomy of job destruction versus job creation to propose a new framework centered on the decentralization of employment. In contrast to the agricultural, industrial, and information ages—where employment was predominantly created by a centralized class of landowners, factory owners, or corporations—the AI era empowers individuals to become creators of their own economic roles. By augmenting human capabilities, AI drastically reduces the scale and capital required to generate economic value, fostering an ecosystem of self-created employment.

This study will proceed as follows. Section 2 reviews the theoretical and empirical literature on technological change and employment, contrasting theories of technological unemployment with the task-based model of labor transformation. Section 3 provides a historical analysis of how previous technological revolutions have consistently expanded the domain of human work. Section 4 elaborates on the concept of an AI Civilization and its role in augmenting human capabilities. Section 5 introduces the core thesis of decentralized employment creation, supported by evidence from the modern creator and platform economies. Section 6 explores the mechanism of human-AI co-evolution as the primary engine of productivity. Section 7 discusses the profound implications for education systems, which must be reoriented to foster creativity and collaborative intelligence. Finally, Section 8 concludes with a summary of the findings and policy recommendations for navigating this historic transformation.

## 2. Technological Change and Employment: A Theoretical Overview

The relationship between technological innovation and employment levels has been a subject of economic debate for over two centuries. The discourse oscillates between fears of mass unemployment driven by automation and theories highlighting the compensatory mechanisms that generate new forms of work. Understanding this theoretical landscape is crucial for contextualizing the unique impact of AI.

## 2.1 Theories of Technological Unemployment

The concern that technology could render human labor obsolete, often termed "technological unemployment," dates back to the classical economists. David Ricardo, in his Principles of Political Economy, famously added a chapter on machinery, admitting that the "substitution of machinery for human labour, is often very injurious to the interests of the class of labourers" in the short term [8]. This perspective was echoed by Karl Marx, who saw mechanization as a tool for capital to increase surplus value at the expense of the workforce.

In the 21st century, this debate has been reignited with greater intensity due to the cognitive capabilities of AI. The most cited study in this vein is by Frey and Osborne (2017), which estimated that approximately 47% of total US employment is at high risk of automation in the coming decades [9]. Their methodology focused on evaluating the susceptibility of entire occupations to computerization based on their core tasks. While influential, this approach has been criticized for potentially overstating the impact by not accounting for the heterogeneity of tasks within a single occupation [6]. Subsequent studies, such as one by Arntz, Gregory, and Zierahn (2016) for the OECD, found that when focusing on the automation potential of individual tasks rather than entire jobs, the share of jobs at high risk drops to a more modest 9% across OECD countries [10].

## 2.2 Task Transformation and the Reinstatement Effect

A more nuanced and widely accepted view in contemporary economics is the task-based framework, most notably developed by Daron Acemoglu and David Autor [6, 7, 11]. This model posits that technology, including AI, does not simply eliminate jobs but rather transforms them by substituting for some tasks while complementing others. The overall impact on employment depends on the balance between two opposing forces: the Displacement Effect, whereby automation directly replaces human labor in specific tasks, and the Reinstatement Effect, whereby technological progress creates new tasks and roles where humans have a comparative advantage. Acemoglu and Restrepo (2020) argue that this reinstatement effect has historically been a powerful countervailing force, as new technologies create new needs, new industries, and new complex tasks that require human skills such as creativity, critical thinking, and social intelligence [2].

For example, while accounting software displaced manual bookkeepers, it simultaneously created a greater demand for financial analysts and auditors who could interpret the data and provide strategic advice. Autor (2015) compellingly asks, "Why are there still so many jobs?" and concludes that it is precisely because technology both destroys and creates work, with the creation effect historically outweighing the destruction [6]. The core challenge of the AI era, therefore, is to ensure that the reinstatement effect is sufficiently strong to offset the powerful displacement driven by AI.

## 3. The Historical Trajectory of Work: An Expanding Frontier

Economic history provides a powerful antidote to technological determinism. Each major technological revolution, while disruptive in the short term, has ultimately led to an expansion of economic activity and a redefinition of human work, rather than a net reduction in employment. This historical pattern is foundational to understanding the potential long-term impact of AI.

## 3.1 From Agricultural to Industrial Civilization

For millennia, human society was organized around agriculture. In these Agricultural Civilizations, land was the primary factor of production, and the vast majority of the population was engaged in farming [12]. Employment was largely static and tied to feudal or land-based hierarchies, with landlords and aristocrats creating the limited employment opportunities available.

The Industrial Revolution, beginning in the late 18th century, represented the first great transformation. The invention of the steam engine, the mechanization of textile production, and the rise of the factory system led to a massive displacement of agricultural labor. This triggered widespread social unrest and the emergence of movements like the Luddites, who feared that machines would make human labor redundant [5]. However, the displacement effect was overwhelmingly surpassed by the reinstatement effect. The factory system created entirely new categories of work—from machine operators and engineers to managers and railway workers. This transition not only absorbed the displaced agricultural workforce but also led to an unprecedented expansion of the economy and a sustained rise in living standards [13].

## 3.2 The Information Civilization and the Service Economy

The late 20th century ushered in the Information Civilization, driven by the microprocessor and the internet. The rise of computers and automation in clerical and manufacturing tasks once again sparked fears of widespread unemployment. The automation of routine tasks, such as those performed by telephone operators, typists, and assembly-line workers, was significant [7]. However, as with the Industrial Revolution, the information age created far more roles than it destroyed. It gave rise to entirely new industries, including software development, information technology services, digital marketing, and data analysis [14]. The digital infrastructure enabled the rapid growth of a global service economy, where value creation shifted from physical goods to information and knowledge.

The table below illustrates how each civilization was built upon the last, with technology consistently serving as a catalyst for expanding the scope of human economic endeavor.

| Civilization | Primary Economic Factor | Dominant Employment Structure | Key Technology | Impact on Work |
|---|---|---|---|---|
| Agricultural | Land | Feudal / Land-based | Plow, Irrigation | Subsistence farming, limited specialization |
| Industrial | Capital / Factories | Centralized (Factory Owners) | Steam Engine, Machinery | Mass displacement from agriculture, rise of factory work |
| Information | Information / Knowledge | Centralized (Corporations) | Computer, Internet | Automation of routine tasks, rise of knowledge and service work |
| AI | Intelligence / Data | Decentralized (Individuals) | AI, Robotics | Augmentation of cognitive tasks, rise of self-created employment |

This historical arc demonstrates that the frontier of human work is not static; it is continuously redefined by technology. The AI revolution should be viewed as the next stage in this evolutionary process. While the displacement effects are real and significant, history suggests that the most profound impact will be the creation of new, currently unimaginable forms of work.

## 4. The AI Civilization: Augmenting and Expanding Human Capability

The AI Civilization represents a distinct phase of technological development because AI systems are not merely tools for automating physical or routine cognitive labor; they are partners in cognitive tasks, capable of augmenting human intelligence, creativity, and problem-solving abilities. Unlike previous technologies that primarily substituted for muscle power or basic calculation, AI is beginning to perform tasks that were once the exclusive domain of human cognition, such as pattern recognition, natural language processing, and complex decision support [15].

However, the narrative of substitution is incomplete. The more profound impact of AI lies in its capacity for human augmentation. By collaborating with AI, individuals can achieve levels of productivity and creativity that were previously unattainable or required the resources of a large organization. This phenomenon is described by Brynjolfsson and Mitchell (2017) as a new form of human-computer symbiosis, where AI handles the complex data analysis and pattern matching, freeing humans to focus on higher-level strategy, creative ideation, and empathetic interaction [16].

Examples of this augmentation are emerging across various fields. In software development, AI-powered tools like GitHub Copilot can write entire blocks of code, allowing developers to build applications faster and focus on system architecture and user experience. In scientific research, AI systems can analyze massive datasets to identify patterns and generate hypotheses, accelerating the pace of discovery in fields from medicine to materials science. In the creative arts, generative AI models enable artists, designers, and musicians to rapidly prototype ideas, create novel visual styles, and produce complex works that would have been prohibitively time-consuming.

This augmentation leads to what this paper terms human-AI co-evolution. As humans become more adept at working with AI, they develop new skills and workflows. In turn, AI systems, trained on the outputs of this collaboration, become even more capable. This virtuous cycle does not lead to a static equilibrium but to a continuous expansion of the frontier of what is possible, creating new economic domains and opportunities [17].

## 5. The Decentralization of Employment Creation

A central thesis of this paper is that the AI Civilization will be characterized by a fundamental shift in the locus of employment creation—from centralized institutions to decentralized individuals. As argued in Section 3, the agricultural, industrial, and information eras were defined by hierarchical employment structures where a small group of capital owners (landlords, factory owners, corporations) created jobs for a large workforce. AI inverts this model.

By providing widespread access to sophisticated cognitive tools, AI dramatically lowers the barriers to entry for creating and delivering value. An individual entrepreneur, artist, or consultant, armed with a suite of AI tools, can now perform the work that once required a team of specialists. This empowers a move towards what we call a self-created employment ecosystem. Evidence for this structural shift can already be observed in the rapid growth of the creator economy and the platform-based gig economy. Millions of individuals are now generating significant income as independent content creators, developers, and freelancers, effectively creating their own jobs outside of traditional corporate structures [18].

Content creation platforms like YouTube, Substack, and TikTok enable individuals to build audiences and monetize their creative output directly, with AI tools for video editing and writing assistance amplifying their productivity. App stores and software marketplaces allow individual developers to build and distribute products to a global audience, with AI-powered coding assistants reducing the complexity of software creation. E-commerce platforms combined with AI-driven marketing tools allow individuals to launch and scale online businesses with minimal upfront investment.

This is not merely an extension of the gig economy, which has often been criticized for its precariousness [19]. While the gig economy unbundled jobs into tasks, the AI-driven creator economy allows for the rebundling of those tasks into sustainable, individual-led businesses. It represents a shift from being a task-taker in a corporate hierarchy to being a value-creator in a networked ecosystem. This decentralization of employment creation is arguably the most significant, yet under-appreciated, socio-economic consequence of the AI revolution.

## 6. Human-AI Co-evolution: The Engine of Future Productivity

The dominant mode of production in the AI Civilization will not be full automation but rather human-AI collaboration. This collaborative paradigm moves beyond a simple master-tool relationship and fosters a dynamic of co-evolution, where the capabilities of both humans and AI systems are mutually enhanced over time. While early examples like AlphaGo demonstrated AI surpassing human experts in narrow domains [20], the broader impact lies in AI augmenting, not replacing, the spectrum of human work.

Recent research provides empirical support for this symbiotic relationship. Studies have shown that human-AI collaboration can lead to significant performance improvements compared to either humans or AI working alone, particularly in complex, creative, and judgmental tasks [21, 22]. A 2024 study in Nature Human Behaviour analyzed various collaborative scenarios and found that while poorly designed human-AI teams could underperform, well-structured collaborations consistently unlocked superior outcomes [23]. The key lies in designing systems that leverage the respective strengths of each partner: the computational power, pattern recognition, and memory of AI, combined with the contextual understanding, ethical judgment, and creative intuition of humans.

This co-evolutionary process unfolds in a feedback loop. First, AI tools augment human workers, allowing them to complete tasks faster, more accurately, and with greater creative capacity. Second, workers adapt their skills and workflows to effectively leverage these new AI capabilities, focusing on higher-value activities that the AI cannot perform. Third, the data generated from these collaborative activities provides feedback for training the next generation of AI models, making them even more capable and useful as collaborators. This dynamic suggests that productivity gains in the AI era will not come from simply replacing workers with algorithms, but from redesigning work processes around the principle of human-AI collaboration—a fundamental shift from automation-as-substitution to an automation-as-augmentation framework.

## 7. Implications for Education: Cultivating Collaborative Intelligence

The structural transformation of work driven by AI necessitates a corresponding revolution in education. Traditional educational models, designed for the industrial and information ages, have emphasized the memorization of facts and the acquisition of procedural knowledge. However, in an era where AI provides instant access to vast stores of information, the value of such skills is diminishing rapidly. The future of education lies not in competing with AI, but in cultivating the uniquely human skills required to collaborate with it.

The Future of Jobs Report 2023 identifies analytical thinking, creative thinking, and AI and big data literacy as the most critical skills for the coming years [4]. This points to a necessary shift in educational focus across three dimensions. First, education must move from knowledge accumulation to problem formulation, teaching students not what to think but how to think—developing skills in critical thinking, formulating insightful questions, and designing complex problem-solving strategies. Second, as AI automates routine tasks, the premium on human creativity will soar, requiring curricula that foster interdisciplinary thinking and encourage students to connect ideas from disparate fields to design novel solutions. Third, the ability to work effectively as part of a human-AI team will become a core competency, requiring training in communication, collaboration, and the ethical and effective use of AI tools.

This requires a systemic overhaul, from primary schools to corporate training programs. Curricula must be made more flexible and project-based, and lifelong learning must become the norm to allow the workforce to adapt to a continuously evolving technological landscape. The goal is to produce a generation of workers who see AI not as a threat, but as a powerful partner in innovation.

## 8. Conclusion: Navigating the Transition to a Decentralized Work Future

The emergence of the AI Civilization is not merely another chapter in the history of technological advancement; it represents a fundamental turning point in the organization of human work. While the narrative of technological unemployment continues to dominate public discourse, this paper has argued that such a view is incomplete. By analyzing the historical trajectory of labor, the task-based nature of automation, and the augmenting power of AI, we have constructed a more optimistic, albeit challenging, vision for the future of work.

The core argument of this paper is that AI, by dramatically enhancing individual productivity, is catalyzing a structural shift from centralized to decentralized employment creation. This transition empowers individuals to become creators of their own economic destinies, fostering a more dynamic and entrepreneurial labor market. The engine of this new economy will be human-AI co-evolution, where the symbiotic partnership between human ingenuity and machine intelligence unlocks unprecedented levels of innovation and value creation.

However, this transition is not automatic. It presents both immense opportunities and significant challenges. The displacement of jobs with high concentrations of routine tasks is inevitable, and the potential for increased inequality is real. Navigating this transformation successfully requires proactive and strategic policy interventions. Governments, businesses, and educational institutions must collaborate to: (1) reform education and workforce development by overhauling curricula to prioritize creative and critical thinking, digital literacy, and the skills for effective human-AI collaboration; (2) support entrepreneurship and self-employment by creating policies that lower barriers for individuals to start and grow their own businesses and by developing robust social safety nets not tied to traditional employment; and (3) promote inclusive and ethical AI by establishing governance frameworks to ensure that AI is developed and deployed in a manner that is fair, transparent, and beneficial to society as a whole.

Rather than fearing an obsolete future, we should focus on building an inclusive and prosperous AI Civilization. The challenge lies not in stopping the tide of technological progress, but in steering it toward a future where augmented human potential is the primary driver of economic growth and shared prosperity.